\documentclass[aps,prl,twocolumn,superscriptaddress,amsmath,amssymb,floatfix]{revtex4-2}

\usepackage[utf8]{inputenc}
\usepackage{graphicx}
\usepackage{xcolor}
\usepackage{bm}
\usepackage{mathtools}
\usepackage{booktabs}

\begin{document}
	
	\title{Optically Active Fractional Wannier-Center Displacement Drives Giant Second-Harmonic Generation}
	
	\author{Hongsheng Pang}
	\affiliation{Laboratory of Quantum Information, University of Science and Technology of China, Hefei 230026, China}
	
	\author{Lixin He}
	\email{helx@ustc.edu.cn}
	\affiliation{Laboratory of Quantum Information, University of Science and Technology of China, Hefei 230026, China}
	\affiliation{Institute of Artificial Intelligence, Hefei Comprehensive National Science Center, Hefei 230088, China}
	\affiliation{Hefei National Laboratory, University of Science and Technology of China, Hefei 230088, China}
	
	\date{\today}
	
	\begin{abstract}
		Electric polarization is a static ground-state Berry-phase property, whereas second-harmonic generation (SHG) and shift current are dynamical optical responses.
		Their connection is encoded in the shift vector, whose Brillouin-zone average is governed by the band-resolved Berry-phase polarization difference between the optically connected initial and final states.
		Here we exploit this geometric relation in quantized formal polarization (QFP) crystals, where symmetry-quantized formal-polarization branches correspond to fractional Wannier-center sectors.
		First-principles screening identifies noncentrosymmetric QFP materials with giant SHG responses, including $\mathrm{InNbBr}_6$ and $\mathrm{InPS}_3$.
		Band-resolved Berry-phase analysis shows that their dominant optical transitions connect occupied and low-lying unoccupied states whose Wannier centers lie at distinct fractional Wyckoff positions, producing a large transition-resolved Wannier-center displacement.
		This displacement gives rise to a large shift vector and a dominant shift-vector-related intraband contribution to the static SHG susceptibility.
	Our results show that symmetry-quantized formal polarization can become optically active through transitions between fractional Wannier-center sectors, providing a symmetry-guided route to giant SHG and shift-current responses.
	\end{abstract}

	\maketitle
	
	{\it Introduction:}
	Second-harmonic generation (SHG) and shift current are prototypical second-order nonlinear optical responses in noncentrosymmetric crystals, with broad relevance to frequency conversion, nonlinear photonics, optoelectronics, and symmetry-resolved material characterization~\cite{shen1984principles,li2021full,ngo2022fibre,fiebig2005second,li2013probing}.
	While inversion-symmetry breaking is required in the electric-dipole approximation, the magnitude of these responses is controlled not only by band gaps and optical transition matrix elements, but also by the geometry of electronic wave functions~\cite{sipe,fregoso2017quantitative}.
	A central geometric quantity is the shift vector, which describes the real-space displacement of an electronic wave packet during an optical transition and enters the intraband contribution to the static SHG susceptibility~\cite{sipe,rashkeev,qian2022role,sturman2021photovoltaic}.
	This suggests a microscopic route to giant nonlinear response: an optically allowed transition with a large real-space displacement between the Wannier centers of the initial and final states can make a large contribution.
	
	This viewpoint provides a natural bridge between nonlinear optics and Berry-phase polarization.
	In the modern theory of polarization~\cite{king1993theory}, the formal electric polarization of an insulator is determined by the positions of occupied electronic Wannier centers relative to the ionic background, modulo a polarization quantum~\cite{resta1994macroscopic,vanderbilt2018berry}.
	By contrast, SHG and shift current are dynamical responses involving optical transitions from occupied to unoccupied states.
	A central question is therefore whether a static Berry-phase polarization structure can become optically active through transitions between distinct Wannier-center sectors.
	
	In certain high-symmetry crystals, crystalline symmetry constrains the formal polarization to discrete values modulo a polarization quantum ~\cite{vanderbilt2018berry,FQFE,pang2025generalized}. We refer to this symmetry-enforced quantization of formal polarization as quantized formal polarization (QFP).
	In a Wannier representation, a fractional QFP corresponds to  symmetry-constrained occupied Wannier centers at fractional Wyckoff positions.
	Previous studies have shown that such QFPs can act as symmetry-protected bulk invariants, constraining band evolution, phase transitions, and interface polar mismatch responses~\cite{pang2025generalized,pang2025quantized,pang2026insulator}.
	
	QFP is a ground-state property and does not by itself guarantee a large nonlinear optical response.
	However, previous studies of QFP-related band evolution~\cite{pang2026insulator} revealed an important electronic-structure motif: occupied and low-lying unoccupied states in certain QFP crystals can belong to distinct symmetry-quantized fractional Wannier-center sectors.
	This observation motivates the present work: if an allowed optical transition connects these sectors, it carries a transition-resolved fractional Wannier-center displacement, which can generate a large shift vector and thereby provide a geometric route to giant SHG and shift-current responses.
The present work therefore establishes a direct link between symmetry-quantized bulk polarization and optically active geometric displacement in nonlinear optical responses.

	Guided by this criterion, we combine first-principles screening with band-resolved Berry-phase analysis to identify noncentrosymmetric QFP materials with giant SHG responses.
	Representative two-dimensional crystals verify this scenario by showing that their large responses originate from optically active transitions between distinct fractional Wannier-center sectors.
	Our results establish transition-resolved fractional Wannier-center displacement as a geometric mechanism for giant SHG and shift-current responses.

	{\it Geometric mechanism:}
	We first show that the intraband contribution to static SHG and the shift-current response share a common geometric factor associated with transition-resolved Wannier-center displacement.
	We focus on the real part of the static SHG susceptibility, $\chi^{(2)}(\omega\rightarrow0)$.
	Following Refs.~\cite{rashkeev,sipe}, the susceptibility can be decomposed into interband and intraband parts,
	\begin{equation}
		\chi^{(2)}=\chi^{(2)}_{\rm inter}+\chi^{(2)}_{\rm intra},
	\end{equation}
	with the full expressions given in the Supplemental Material (SM)~\cite{SM}.
	
	The interband contribution is built from products of interband optical matrix elements $r^a_{mn}$, whereas the intraband contribution contains their covariant $k$-derivatives and thus explicitly encodes the wave-function geometry.
	It therefore provides a direct channel for enhancing SHG through a large Wannier-center displacement.
	
	For  $r^a_{mn}=|r^a_{mn}|e^{i\phi^a_{mn}}$, the generalized derivative is
\begin{equation}
r^a_{mn;b}=\partial_{k_b}r^a_{mn} - i\left(A^b_{mm}-A^b_{nn}\right)r^a_{mn},
	\end{equation}
	where $A^b_{nn}$ is the intraband Berry connection of band $n$.
	
	For an optical transition $n\rightarrow m$, the shift vector is defined as~\cite{sturman2021photovoltaic,sipe}
\begin{equation}
R^a_{mn;b} = \partial_{k_b}\phi^a_{mn} - A^b_{mm} + A^b_{nn},
	\end{equation}
	which is a gauge-invariant measure of the real-space displacement of the electronic wave packet during the transition.
	Combining this definition with the generalized derivative gives
	\begin{equation}
		{\rm Im}\left[ r^a_{nm}r^a_{mn;b} \right] =|r^a_{mn}|^2R^a_{mn;b}.
		\label{eq:shift_factor}
	\end{equation}
	The real part of the static intraband SHG susceptibility therefore contains terms proportional to the shift-vector-related geometric factor $|r|^2R$, up to conventional signs and tensor permutations~\cite{SM}.
	
	The same geometric factor appears directly in the shift-current response~\cite{sipe,jin2024peculiar,young2012first},
	\begin{equation}
		\sigma^{abb}(\omega)
		\propto
		\int[d\mathbf{k}]
		\sum_{nm}
		f_{nm}|r^b_{nm}|^2
		R^a_{mn;b}
		\delta(\omega_{mn}-\omega),
	\end{equation}
	up to conventional prefactors.
	Thus, both the shift-current response and the intraband part of the static SHG susceptibility are favored by optically allowed transitions that combine sizable optical matrix elements with large shift vectors.
	
	The link to Wannier-center physics follows from the Brillouin-zone integral of the shift vector.
	For isolated initial and final bands~\cite{fregoso2017quantitative},
\begin{equation}
-e\int[d\bm{k}]\,\bm{R}_{mn}	=	\bm{Q}W_{mn}+ \bm{P}_{m}
		- \bm{P}_{n},
		\label{eq:shift_integral}
	\end{equation}
	where $\bm P_n=-e\int[d\bm{k}]\,\bm A_{nn}$ and $\bm P_m=-e\int[d\bm{k}]\,\bm A_{mm}$ are band-resolved Berry-phase polarizations, equivalently Wannier centers, of the initial and final bands, $\bm Q$ is a polarization quantum, and $W_{mn}$ is an integer winding number associated with optical zeros.
When optical zeros are absent, the winding term does not contribute under a proper optical gauge.
	Equation~(\ref{eq:shift_integral}) shows that a large Brillouin-zone-averaged shift vector can arise from a large Berry-phase polarization difference between the occupied initial state and the unoccupied final state.
	This relation motivates the central criterion used here: large second-order nonlinear responses are favored by optically allowed transitions that combine sizable optical matrix elements with distinct initial- and final-state Wannier centers.

	QFP materials provide a natural platform for realizing this criterion.
	Although QFP is a ground-state property of the occupied manifold, QFP-related band structures can host low-lying unoccupied states in symmetry-distinct fractional Wannier-center sectors~\cite{pang2026insulator}.
	In certain QFP materials, optical transitions between these sectors generate a large shift vector through their fractional Berry-phase polarization difference, thereby enhancing shift current and the intraband contribution to static SHG.
	This provides a geometric mechanism for giant second-order nonlinear optical responses.
	
	{\it Materials screening:}
	Guided by this criterion, we use high-symmetry two-dimensional semiconductors as a discovery platform for QFP materials with giant SHG responses.
	Two-dimensional materials are well suited for this search because their atomic-scale thickness, compatibility with photonic integration, and high tunability make them attractive for nanoscale nonlinear optics~\cite{huang2024second,zhang2020second}.
	We screen noncentrosymmetric hexagonal materials from the Computational 2D Materials Database (C2DB)~\cite{haastrup2018computational,gjerding2021recent}, focusing on nonmagnetic insulators whose band gaps calculated with the Heyd--Scuseria--Ernzerhof (HSE) hybrid functional~\cite{hse} are larger than 1.0 eV, in order to avoid trivially enhanced responses from nearly gapless systems.
	We further retain only point groups that satisfy two symmetry requirements: they allow nonzero in-plane second-order nonlinear optical responses and can support symmetry-quantized formal-polarization branches.
	The resulting point groups are $C_3$, $D_3$, $C_{3v}$, $C_{3h}$, and $D_{3h}$.
	Their symmetry-allowed QFP branches and independent in-plane second-order tensor components are listed in the SM~\cite{SM}.
	
	We perform self-consistent density-functional-theory (DFT) calculations using the ABACUS package~\cite{li2016large,chen2010systematically,lin2024abinitio,lin2021strategy}, with the HSE hybrid functional~\cite{hse,lin2020accuracy} and spin-orbit coupling included.
	The resulting tight-binding Hamiltonians are then used as input to the PYATB package~\cite{jin2023} to evaluate the Berry-phase polarization and second-order optical responses, including the shift-current conductivity and SHG susceptibility.
	Further computational details are provided in the SM~\cite{SM}.

	\begin{figure}[t]
		\centering
		\includegraphics[width=0.4\textwidth]{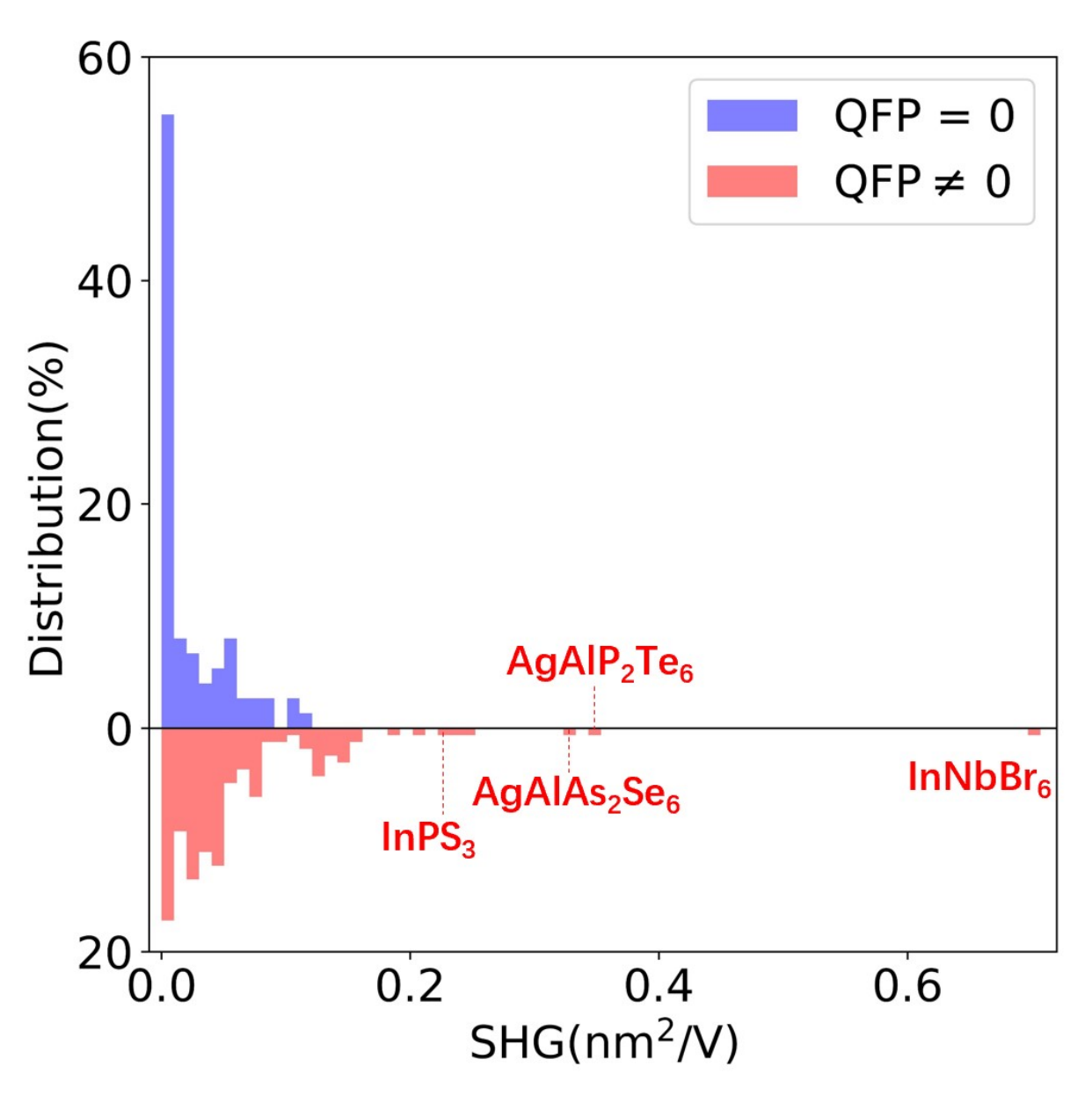}
		\caption{
			Distribution of the HSE+SOC calculated static SHG susceptibility for screened zero-QFP and nonzero-QFP materials.
			Several candidates with exceptionally large static SHG responses are highlighted.
		}
		\label{fig:distribution}
	\end{figure}

	The screening identifies 74 zero-QFP materials and 161 nonzero-QFP materials.
	As shown in Fig.~\ref{fig:distribution}, although QFP is not by itself a sufficient condition for large SHG, the nonzero-QFP group contains a higher fraction of large-response materials than the zero-QFP group.
	In particular, several of the largest-SHG candidates belong to the nonzero-QFP class, including InNbBr$_6$, AgAlAs$_2$Se$_6$, and InPS$_3$, which exhibit HSE+SOC static SHG susceptibilities of 0.70, 0.33, and 0.23 nm$^2$/V, respectively.
	
	Several two-dimensional materials, such as group-IV monochalcogenides in the MX family, have been predicted to exhibit very large SHG responses at the Perdew--Burke--Ernzerhof (PBE) level~\cite{wang2017giant}.
	These values provide a useful scale for comparison, but should not be directly benchmarked against the HSE+SOC results reported here, because semilocal functionals such as PBE typically underestimate band gaps and can overestimate static SHG magnitudes.
	For InNbBr$_6$, a PBE-level calculation reduces the band gap to 0.92 eV and gives $\chi^{(2)}=1.18$ nm$^2$/V, compared with the HSE+SOC value of 0.70 nm$^2$/V.
	Thus, our HSE+SOC results provide a conservative estimate while still placing the identified QFP materials among strong two-dimensional nonlinear optical candidates.
	
	These screening results motivate a band-resolved analysis of whether the large responses originate from optically active transitions between distinct fractional Wannier-center sectors.

\begin{figure}[tb]
		\centering
		\includegraphics[width=0.48\textwidth]{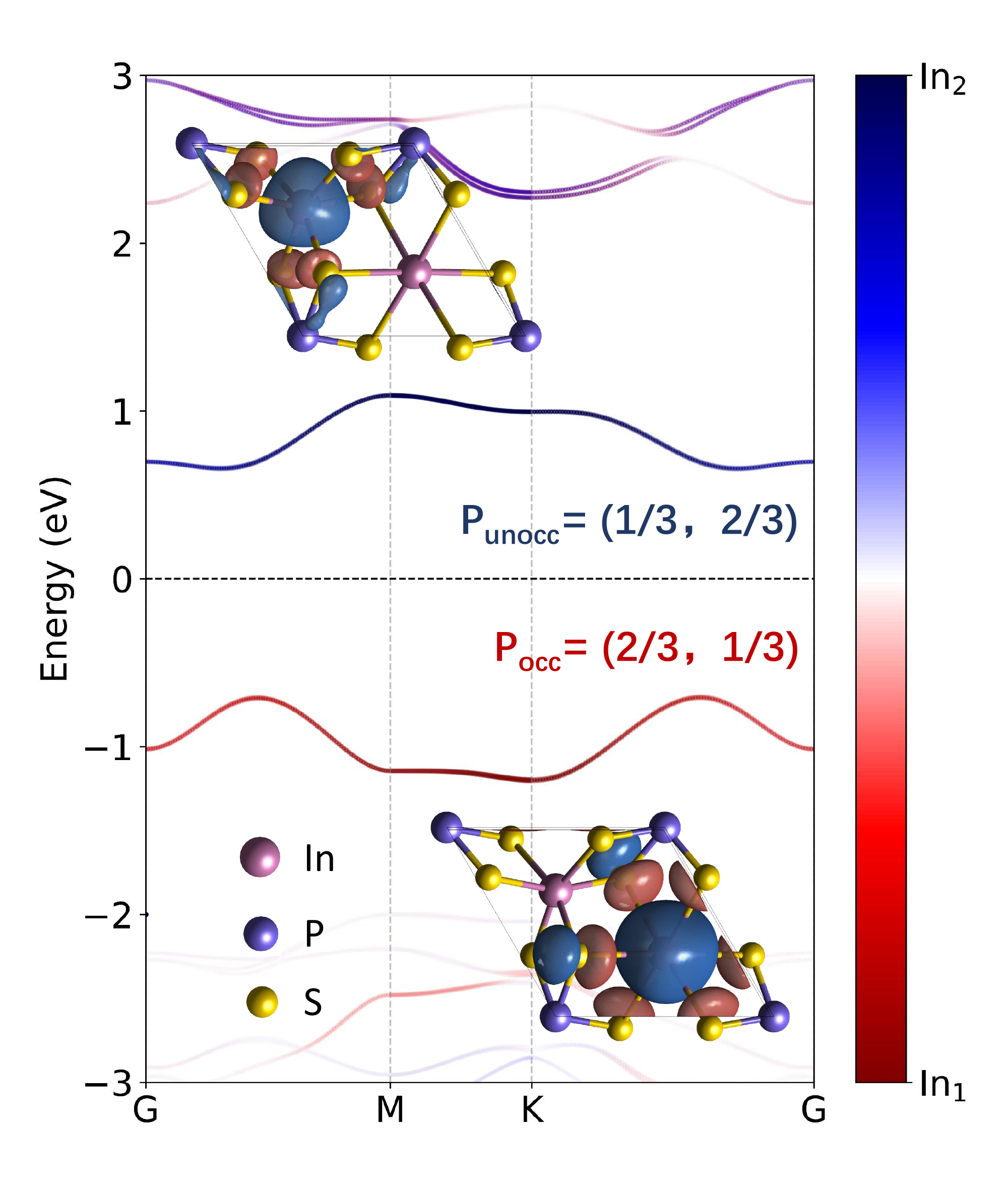}
		\caption{
			Band-edge states and fractional Wannier centers in InPS$_3$.
			The highest occupied and lowest unoccupied bands are mainly localized on the symmetry-inequivalent In$_1$ and In$_2$ sites, respectively, as shown by the fat-band projection and Wannier functions.
		}
		\label{fig:inps3_band}
	\end{figure}

	{\it InPS$_3$:}
	We first examine InPS$_3$.
	Although it does not exhibit the largest SHG response among the screened materials, it provides a particularly clean two-dimensional example of the proposed mechanism.
	InPS$_3$ crystallizes in space group $P312$ and belongs to point group $D_3$.
	Its crystal structure is shown in the inset of Fig.~\ref{fig:inps3_band}.
	The two inequivalent In atoms occupy the fractional Wyckoff positions $(1/3,2/3)$ and $(2/3,1/3)$.
	First-principles calculations give an in-plane formal polarization of $(2/3,1/3)$, consistent with the symmetry-allowed QFP branch~\cite{pang2025generalized}.
	The HSE+SOC calculation yields a band gap of 1.44 eV.
	The highest occupied and lowest unoccupied bands, which are isolated from the other bands near the Fermi level, are shown in Fig.~\ref{fig:inps3_band}.

	\begin{figure}[t]
		\centering
		\includegraphics[width=0.4\textwidth]{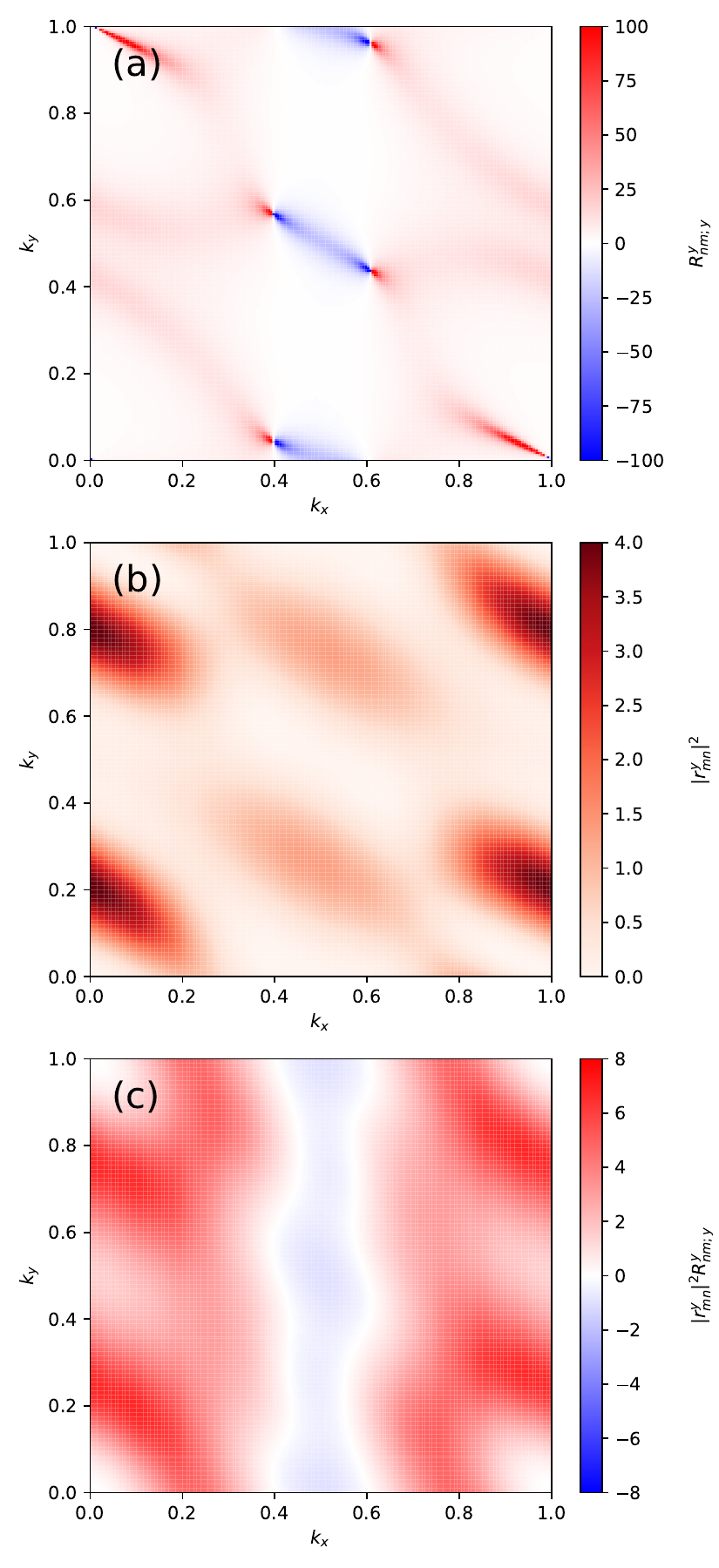}
		\caption{
			Momentum-resolved geometric quantities for the band-edge transition in InPS$_3$:
			(a) shift vector $R$,
			(b) optical transition strength $|r|^2$, and
			(c) shift-vector-weighted optical factor $|r|^2R$.
		}
		\label{fig:inps3_shift_vector}
	\end{figure}
	
	We decompose the static SHG susceptibility of InPS$_3$ into interband and intraband contributions.
	The total static susceptibility is $0.23$ nm$^2$/V, with intraband and interband contributions of $0.24$ and $-0.01$ nm$^2$/V, respectively.
	Thus, the intraband term nearly reproduces the total response, whereas the interband contribution is negligible and slightly negative.
	This decomposition makes InPS$_3$ a clean example in which the static SHG response is governed primarily by the shift-vector-related intraband channel.
	
	The frequency-dependent responses further identify the relevant low-energy optical transition.
	As shown in Fig.~S1 of the SM~\cite{SM}, the shift-current conductivity exhibits a pronounced feature near the band gap, while the SHG susceptibility shows a strong half-gap resonance.
	These spectral features indicate that the band-edge transition between the highest occupied and lowest unoccupied bands provides the dominant low-energy channel relevant to the nonlinear response.
	We therefore trace the microscopic origin of the intraband response to the real-space character of these band-edge states.
	
	To this end, we construct Wannier functions separately for the isolated highest occupied and lowest unoccupied bands.
	As shown in Fig.~\ref{fig:inps3_band}, the Wannier center of the highest occupied band is centered on the In$_1$ site at the fractional Wyckoff position $(2/3,1/3)$, whereas that of the lowest unoccupied band is centered on the symmetry-inequivalent In$_2$ site at $(1/3,2/3)$.
	The corresponding band-resolved Berry-phase polarizations are
	\begin{equation}
		\bm{P}_{\rm occ}=(2/3,1/3), \qquad
		\bm{P}_{\rm unocc}=(1/3,2/3).
	\end{equation}
	Thus, the band-edge optical transition connects two distinct fractional Wannier-center sectors.
	According to Eq.~(\ref{eq:shift_integral}), this transition-resolved Berry-phase polarization difference provides the geometric origin of the large Brillouin-zone-averaged shift vector.
	
	We further examine this mechanism in momentum space.
	Figure~\ref{fig:inps3_shift_vector}(a) shows that the band-edge transition produces large shift vectors over extended regions of the Brillouin zone, consistent with the fractional Wannier-center displacement identified above.
	To verify that this geometric displacement is optically active, we evaluate the optical transition strength $|r|^2$ and the shift-vector-weighted factor $|r|^2R$ [Figs.~\ref{fig:inps3_shift_vector}(b,c)].
	The pronounced $|r|^2R$ distribution shows that the large shift vector is accompanied by sizable optical matrix elements and therefore feeds directly into the dominant intraband SHG response.

	\begin{figure}[t]
		\centering
		\includegraphics[width=0.4\textwidth]{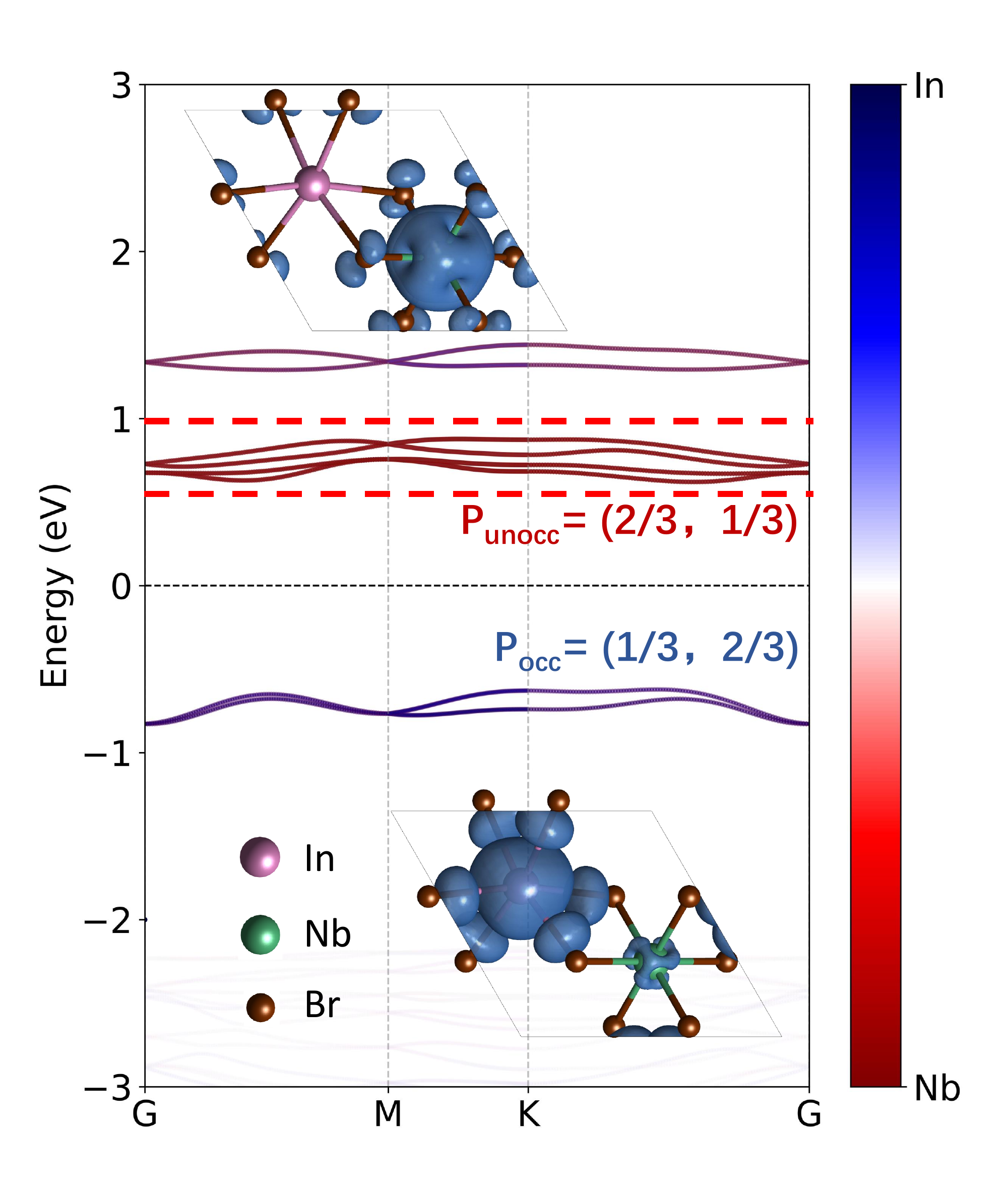}
		\caption{
			Projected band structure and partial charge densities of InNbBr$_6$.
			The occupied band-edge states are mainly localized on the In sublattice, whereas the selected low-lying unoccupied states  are mainly localized on the Nb sublattice.
			Insets show the summed partial charge densities of the occupied and unoccupied states.
			The unoccupied-state charge density is integrated over the 4 states enclosed by the two red dashed boxes, while the occupied-state charge density is obtained from the two occupied bands closest to the Fermi level.
		}
		\label{fig:innbbr6_band}
	\end{figure}

	{\it InNbBr$_6$:}
	We next consider InNbBr$_6$, the material with the largest static SHG response among the screened candidates.
	It has the same space group $P312$ and point group $D_3$ as InPS$_3$, but exhibits a substantially larger static SHG susceptibility.
	The In and Nb atoms occupy the fractional Wyckoff positions $(1/3,2/3)$ and $(2/3,1/3)$, respectively, and the calculated in-plane formal polarization is $(2/3,1/3)$, consistent with the symmetry-allowed QFP branch.
	The HSE+SOC calculation yields a band gap of 1.24 eV.
	
	We first decompose the static SHG response of InNbBr$_6$ into intraband and interband contributions.
	The intraband term is strongly positive, with a value of $1.44$ nm$^2$/V, whereas the interband term is negative, with a value of $-0.74$ nm$^2$/V, yielding a partially canceled total susceptibility of $0.70$ nm$^2$/V. Despite this substantial cancellation, InNbBr$_6$ exhibits the largest response among the screened materials, indicating that its SHG response is primarily governed by the shift-vector-related intraband channel.
	
	The frequency-dependent spectra identify the relevant low-energy transitions.
	As shown in Fig.~S2 in the SM~\cite{SM}, the SHG susceptibility exhibits strong resonant features associated with the band-edge transitions, while the shift-current conductivity shows a pronounced response near the band gap.
	These spectral features indicate that the low-lying unoccupied states and the occupied band-edge states are the relevant optical channels for the nonlinear response.
	
	We then analyze the real-space character of these band-edge manifolds.
	Near the Fermi level, the relevant unoccupied band-edge states are entangled and involve multiple Wannier centers.
	Therefore, instead of using individual Wannier functions, we characterize the real-space distributions of the occupied and unoccupied band-edge manifolds by their summed partial charge densities.
	The unoccupied-state charge density is summed over the selected low-lying bands marked in Fig.~\ref{fig:innbbr6_band}, while the occupied-state charge density is obtained from the occupied bands closest to the Fermi level.
	As shown in Fig.~\ref{fig:innbbr6_band}, the occupied band-edge states are mainly localized on the In sublattice, whereas the selected low-lying unoccupied states are mainly localized on the Nb sublattice.
	Their charge distributions are centered around distinct fractional positions in the unit cell.
	The corresponding Berry-phase polarizations of the selected occupied and unoccupied band-edge manifolds are
	\begin{equation}
		\bm{P}_{\rm occ}=(1/3,2/3), \qquad
		\bm{P}_{\rm unocc}=(2/3,1/3).
	\end{equation}
	Thus, as in InPS$_3$, the relevant low-energy optical transitions connect distinct fractional Wannier-center sectors, providing the Berry-phase origin of a large Brillouin-zone-averaged shift vector according to Eq.~(\ref{eq:shift_integral}).
	
	The $\bm{k}$-resolved distributions in Fig.~S4 further show that the shift-vector-weighted response is spread over extended regions of the Brillouin zone, confirming that the fractional Wannier-center displacement is optically active in InNbBr$_6$.

Although the main text focuses on two-dimensional materials, the mechanism is not restricted to two-dimensional systems.
Since the underlying principle relies only on a Berry-phase polarization difference between optically connected states, it should also apply to three-dimensional QFP crystals.
In the SM, we present CdBiO$_3$ as a representative three-dimensional example that exhibits a significant SHG response through the same transition-resolved fractional Wannier-center mechanism~\cite{SM}.

{\it Summary:}
We identify transition-resolved fractional Wannier-center displacement as a geometric mechanism for giant second-order nonlinear optical responses.
In QFP materials, optically allowed transitions between occupied and low-lying unoccupied states in distinct fractional Wannier-center sectors can carry a lattice-scale displacement, generating a large shift vector and a dominant shift-vector-related intraband contribution to the static SHG susceptibility.
First-principles screening and representative material analyses reveal such optically active transitions in noncentrosymmetric QFP materials with giant SHG responses and pronounced shift-current responses.
These results show that symmetry-quantized formal polarization, although a static bulk property, can become an optically active geometric ingredient in dynamical nonlinear responses.
They establish a symmetry-guided design principle for high-performance nonlinear optical materials based on optical transitions between distinct fractional Wannier-center sectors.
	
This work was supported by the Advanced Materials–National Science and Technology Major Project (Grant No. 2025ZD0618401), the National Natural Science Foundation of China (Grant No. 12134012), the Strategic Priority Research Program of the Chinese Academy of Sciences (Grant No. XDB0500201), and the Innovation Program for Quantum Science and Technology (Grant No. 2021ZD0301200). The numerical calculations were performed on the USTC High-Performance Computing facilities.	
	
%\bibliography{mybib,abacus_refs}

\begin{thebibliography}{38}%
\makeatletter
\providecommand \@ifxundefined [1]{%
 \@ifx{#1\undefined}
}%
\providecommand \@ifnum [1]{%
 \ifnum #1\expandafter \@firstoftwo
 \else \expandafter \@secondoftwo
 \fi
}%
\providecommand \@ifx [1]{%
 \ifx #1\expandafter \@firstoftwo
 \else \expandafter \@secondoftwo
 \fi
}%
\providecommand \natexlab [1]{#1}%
\providecommand \enquote  [1]{``#1''}%
\providecommand \bibnamefont  [1]{#1}%
\providecommand \bibfnamefont [1]{#1}%
\providecommand \citenamefont [1]{#1}%
\providecommand \href@noop [0]{\@secondoftwo}%
\providecommand \href [0]{\begingroup \@sanitize@url \@href}%
\providecommand \@href[1]{\@@startlink{#1}\@@href}%
\providecommand \@@href[1]{\endgroup#1\@@endlink}%
\providecommand \@sanitize@url [0]{\catcode `\\12\catcode `\$12\catcode
  `\&12\catcode `\#12\catcode `\^12\catcode `\_12\catcode `\%12\relax}%
\providecommand \@@startlink[1]{}%
\providecommand \@@endlink[0]{}%
\providecommand \url  [0]{\begingroup\@sanitize@url \@url }%
\providecommand \@url [1]{\endgroup\@href {#1}{\urlprefix }}%
\providecommand \urlprefix  [0]{URL }%
\providecommand \Eprint [0]{\href }%
\providecommand \doibase [0]{https://doi.org/}%
\providecommand \selectlanguage [0]{\@gobble}%
\providecommand \bibinfo  [0]{\@secondoftwo}%
\providecommand \bibfield  [0]{\@secondoftwo}%
\providecommand \translation [1]{[#1]}%
\providecommand \BibitemOpen [0]{}%
\providecommand \bibitemStop [0]{}%
\providecommand \bibitemNoStop [0]{.\EOS\space}%
\providecommand \EOS [0]{\spacefactor3000\relax}%
\providecommand \BibitemShut  [1]{\csname bibitem#1\endcsname}%
\let\auto@bib@innerbib\@empty
%</preamble>
\bibitem [{\citenamefont {Shen}(1984)}]{shen1984principles}%
  \BibitemOpen
  \bibfield  {author} {\bibinfo {author} {\bibfnamefont {Y.~R.}\ \bibnamefont
  {Shen}},\ }\href@noop {} {\emph {\bibinfo {title} {The Principles of
  Nonlinear Optics}}}\ (\bibinfo  {publisher} {John Wiley \& Sons},\ \bibinfo
  {address} {New York},\ \bibinfo {year} {1984})\BibitemShut {NoStop}%
\bibitem [{\citenamefont {Li}\ \emph {et~al.}(2021)\citenamefont {Li},
  \citenamefont {Hu}, \citenamefont {Shi}, \citenamefont {He}, \citenamefont
  {Li}, \citenamefont {Shang}, \citenamefont {Zhang}, \citenamefont {Fu},
  \citenamefont {Zhou}, \citenamefont {Xiong} \emph {et~al.}}]{li2021full}%
  \BibitemOpen
  \bibfield  {author} {\bibinfo {author} {\bibfnamefont {J.}~\bibnamefont
  {Li}}, \bibinfo {author} {\bibfnamefont {G.}~\bibnamefont {Hu}}, \bibinfo
  {author} {\bibfnamefont {L.}~\bibnamefont {Shi}}, \bibinfo {author}
  {\bibfnamefont {N.}~\bibnamefont {He}}, \bibinfo {author} {\bibfnamefont
  {D.}~\bibnamefont {Li}}, \bibinfo {author} {\bibfnamefont {Q.}~\bibnamefont
  {Shang}}, \bibinfo {author} {\bibfnamefont {Q.}~\bibnamefont {Zhang}},
  \bibinfo {author} {\bibfnamefont {H.}~\bibnamefont {Fu}}, \bibinfo {author}
  {\bibfnamefont {L.}~\bibnamefont {Zhou}}, \bibinfo {author} {\bibfnamefont
  {W.}~\bibnamefont {Xiong}}, \emph {et~al.},\ }\bibfield  {title} {\bibinfo
  {title} {Full-color enhanced second harmonic generation using rainbow
  trapping in ultrathin hyperbolic metamaterials},\ }\href@noop {} {\bibfield
  {journal} {\bibinfo  {journal} {Nat. Commun.}\ }\textbf {\bibinfo {volume}
  {12}},\ \bibinfo {pages} {6425} (\bibinfo {year} {2021})}\BibitemShut
  {NoStop}%
\bibitem [{\citenamefont {Ngo}\ \emph {et~al.}(2022)\citenamefont {Ngo},
  \citenamefont {Najafidehaghani}, \citenamefont {Gan}, \citenamefont
  {Khazaee}, \citenamefont {Siems}, \citenamefont {George}, \citenamefont
  {Schartner}, \citenamefont {Nolte}, \citenamefont {Ebendorff-Heidepriem},
  \citenamefont {Pertsch} \emph {et~al.}}]{ngo2022fibre}%
  \BibitemOpen
  \bibfield  {author} {\bibinfo {author} {\bibfnamefont {G.~Q.}\ \bibnamefont
  {Ngo}}, \bibinfo {author} {\bibfnamefont {E.}~\bibnamefont
  {Najafidehaghani}}, \bibinfo {author} {\bibfnamefont {Z.}~\bibnamefont
  {Gan}}, \bibinfo {author} {\bibfnamefont {S.}~\bibnamefont {Khazaee}},
  \bibinfo {author} {\bibfnamefont {M.~P.}\ \bibnamefont {Siems}}, \bibinfo
  {author} {\bibfnamefont {A.}~\bibnamefont {George}}, \bibinfo {author}
  {\bibfnamefont {E.~P.}\ \bibnamefont {Schartner}}, \bibinfo {author}
  {\bibfnamefont {S.}~\bibnamefont {Nolte}}, \bibinfo {author} {\bibfnamefont
  {H.}~\bibnamefont {Ebendorff-Heidepriem}}, \bibinfo {author} {\bibfnamefont
  {T.}~\bibnamefont {Pertsch}}, \emph {et~al.},\ }\bibfield  {title} {\bibinfo
  {title} {In-fibre second-harmonic generation with embedded two-dimensional
  materials},\ }\href@noop {} {\bibfield  {journal} {\bibinfo  {journal} {Nat.
  Photonics}\ }\textbf {\bibinfo {volume} {16}},\ \bibinfo {pages} {769}
  (\bibinfo {year} {2022})}\BibitemShut {NoStop}%
\bibitem [{\citenamefont {Fiebig}\ \emph {et~al.}(2005)\citenamefont {Fiebig},
  \citenamefont {Pavlov},\ and\ \citenamefont {Pisarev}}]{fiebig2005second}%
  \BibitemOpen
  \bibfield  {author} {\bibinfo {author} {\bibfnamefont {M.}~\bibnamefont
  {Fiebig}}, \bibinfo {author} {\bibfnamefont {V.~V.}\ \bibnamefont {Pavlov}},\
  and\ \bibinfo {author} {\bibfnamefont {R.~V.}\ \bibnamefont {Pisarev}},\
  }\bibfield  {title} {\bibinfo {title} {Second-harmonic generation as a tool
  for studying electronic and magnetic structures of crystals},\ }\href@noop {}
  {\bibfield  {journal} {\bibinfo  {journal} {J. Opt. Soc. Am. B}\ }\textbf
  {\bibinfo {volume} {22}},\ \bibinfo {pages} {96} (\bibinfo {year}
  {2005})}\BibitemShut {NoStop}%
\bibitem [{\citenamefont {Li}\ \emph {et~al.}(2013)\citenamefont {Li},
  \citenamefont {Rao}, \citenamefont {Mak}, \citenamefont {You}, \citenamefont
  {Wang}, \citenamefont {Dean},\ and\ \citenamefont {Heinz}}]{li2013probing}%
  \BibitemOpen
  \bibfield  {author} {\bibinfo {author} {\bibfnamefont {Y.}~\bibnamefont
  {Li}}, \bibinfo {author} {\bibfnamefont {Y.}~\bibnamefont {Rao}}, \bibinfo
  {author} {\bibfnamefont {K.~F.}\ \bibnamefont {Mak}}, \bibinfo {author}
  {\bibfnamefont {Y.}~\bibnamefont {You}}, \bibinfo {author} {\bibfnamefont
  {S.}~\bibnamefont {Wang}}, \bibinfo {author} {\bibfnamefont {C.~R.}\
  \bibnamefont {Dean}},\ and\ \bibinfo {author} {\bibfnamefont {T.~F.}\
  \bibnamefont {Heinz}},\ }\bibfield  {title} {\bibinfo {title} {Probing
  symmetry properties of few-layer mos2 and h-bn by optical second-harmonic
  generation},\ }\href@noop {} {\bibfield  {journal} {\bibinfo  {journal} {Nano
  Lett.}\ }\textbf {\bibinfo {volume} {13}},\ \bibinfo {pages} {3329} (\bibinfo
  {year} {2013})}\BibitemShut {NoStop}%
\bibitem [{\citenamefont {Sipe}\ and\ \citenamefont {Shkrebtii}(2000)}]{sipe}%
  \BibitemOpen
  \bibfield  {author} {\bibinfo {author} {\bibfnamefont {J.~E.}\ \bibnamefont
  {Sipe}}\ and\ \bibinfo {author} {\bibfnamefont {A.~I.}\ \bibnamefont
  {Shkrebtii}},\ }\bibfield  {title} {\bibinfo {title} {Second-order optical
  response in semiconductors},\ }\href
  {https://doi.org/10.1103/PhysRevB.61.5337} {\bibfield  {journal} {\bibinfo
  {journal} {Phys. Rev. B}\ }\textbf {\bibinfo {volume} {61}},\ \bibinfo
  {pages} {5337} (\bibinfo {year} {2000})}\BibitemShut {NoStop}%
\bibitem [{\citenamefont {Fregoso}\ \emph {et~al.}(2017)\citenamefont
  {Fregoso}, \citenamefont {Morimoto},\ and\ \citenamefont
  {Moore}}]{fregoso2017quantitative}%
  \BibitemOpen
  \bibfield  {author} {\bibinfo {author} {\bibfnamefont {B.~M.}\ \bibnamefont
  {Fregoso}}, \bibinfo {author} {\bibfnamefont {T.}~\bibnamefont {Morimoto}},\
  and\ \bibinfo {author} {\bibfnamefont {J.~E.}\ \bibnamefont {Moore}},\
  }\bibfield  {title} {\bibinfo {title} {Quantitative relationship between
  polarization differences and the zone-averaged shift photocurrent},\
  }\href@noop {} {\bibfield  {journal} {\bibinfo  {journal} {Phys. Rev. B}\
  }\textbf {\bibinfo {volume} {96}},\ \bibinfo {pages} {075421} (\bibinfo
  {year} {2017})}\BibitemShut {NoStop}%
\bibitem [{\citenamefont {Rashkeev}\ \emph {et~al.}(1998)\citenamefont
  {Rashkeev}, \citenamefont {Lambrecht},\ and\ \citenamefont
  {Segall}}]{rashkeev}%
  \BibitemOpen
  \bibfield  {author} {\bibinfo {author} {\bibfnamefont {S.~N.}\ \bibnamefont
  {Rashkeev}}, \bibinfo {author} {\bibfnamefont {W.~R.}\ \bibnamefont
  {Lambrecht}},\ and\ \bibinfo {author} {\bibfnamefont {B.}~\bibnamefont
  {Segall}},\ }\bibfield  {title} {\bibinfo {title} {Efficient ab initio method
  for the calculation of frequency-dependent second-order optical response in
  semiconductors},\ }\href@noop {} {\bibfield  {journal} {\bibinfo  {journal}
  {Phys. Rev. B}\ }\textbf {\bibinfo {volume} {57}},\ \bibinfo {pages} {3905}
  (\bibinfo {year} {1998})}\BibitemShut {NoStop}%
\bibitem [{\citenamefont {Qian}\ \emph {et~al.}(2022)\citenamefont {Qian},
  \citenamefont {Yu}, \citenamefont {Jiang}, \citenamefont {Zhang},
  \citenamefont {Gao}, \citenamefont {Shi}, \citenamefont {Pi}, \citenamefont
  {Weng},\ and\ \citenamefont {Lu}}]{qian2022role}%
  \BibitemOpen
  \bibfield  {author} {\bibinfo {author} {\bibfnamefont {C.}~\bibnamefont
  {Qian}}, \bibinfo {author} {\bibfnamefont {C.}~\bibnamefont {Yu}}, \bibinfo
  {author} {\bibfnamefont {S.}~\bibnamefont {Jiang}}, \bibinfo {author}
  {\bibfnamefont {T.}~\bibnamefont {Zhang}}, \bibinfo {author} {\bibfnamefont
  {J.}~\bibnamefont {Gao}}, \bibinfo {author} {\bibfnamefont {S.}~\bibnamefont
  {Shi}}, \bibinfo {author} {\bibfnamefont {H.}~\bibnamefont {Pi}}, \bibinfo
  {author} {\bibfnamefont {H.}~\bibnamefont {Weng}},\ and\ \bibinfo {author}
  {\bibfnamefont {R.}~\bibnamefont {Lu}},\ }\bibfield  {title} {\bibinfo
  {title} {Role of shift vector in high-harmonic generation from
  noncentrosymmetric topological insulators under strong laser fields},\
  }\href@noop {} {\bibfield  {journal} {\bibinfo  {journal} {Phys. Rev. X}\
  }\textbf {\bibinfo {volume} {12}},\ \bibinfo {pages} {021030} (\bibinfo
  {year} {2022})}\BibitemShut {NoStop}%
\bibitem [{\citenamefont {Sturman}\ and\ \citenamefont
  {Fridkin}(2021)}]{sturman2021photovoltaic}%
  \BibitemOpen
  \bibfield  {author} {\bibinfo {author} {\bibfnamefont {B.}~\bibnamefont
  {Sturman}}\ and\ \bibinfo {author} {\bibfnamefont {V.}~\bibnamefont
  {Fridkin}},\ }\href@noop {} {\emph {\bibinfo {title} {Photovoltaic and
  photo-refractive effects in noncentrosymmetric materials}}}\ (\bibinfo
  {publisher} {Routledge},\ \bibinfo {year} {2021})\BibitemShut {NoStop}%
\bibitem [{\citenamefont {King-Smith}\ and\ \citenamefont
  {Vanderbilt}(1993)}]{king1993theory}%
  \BibitemOpen
  \bibfield  {author} {\bibinfo {author} {\bibfnamefont {R.}~\bibnamefont
  {King-Smith}}\ and\ \bibinfo {author} {\bibfnamefont {D.}~\bibnamefont
  {Vanderbilt}},\ }\bibfield  {title} {\bibinfo {title} {Theory of polarization
  of crystalline solids},\ }\href@noop {} {\bibfield  {journal} {\bibinfo
  {journal} {Phys. Rev. B}\ }\textbf {\bibinfo {volume} {47}},\ \bibinfo
  {pages} {1651} (\bibinfo {year} {1993})}\BibitemShut {NoStop}%
\bibitem [{\citenamefont {Resta}(1994)}]{resta1994macroscopic}%
  \BibitemOpen
  \bibfield  {author} {\bibinfo {author} {\bibfnamefont {R.}~\bibnamefont
  {Resta}},\ }\bibfield  {title} {\bibinfo {title} {Macroscopic polarization in
  crystalline dielectrics: the geometric phase approach},\ }\href@noop {}
  {\bibfield  {journal} {\bibinfo  {journal} {Rev. Mod. Phys.}\ }\textbf
  {\bibinfo {volume} {66}},\ \bibinfo {pages} {899} (\bibinfo {year}
  {1994})}\BibitemShut {NoStop}%
\bibitem [{\citenamefont {Vanderbilt}(2018)}]{vanderbilt2018berry}%
  \BibitemOpen
  \bibfield  {author} {\bibinfo {author} {\bibfnamefont {D.}~\bibnamefont
  {Vanderbilt}},\ }\href@noop {} {\emph {\bibinfo {title} {Berry phases in
  electronic structure theory: electric polarization, orbital magnetization and
  topological insulators}}}\ (\bibinfo  {publisher} {Cambridge University
  Press},\ \bibinfo {address} {Cambridge, UK},\ \bibinfo {year}
  {2018})\BibitemShut {NoStop}%
\bibitem [{\citenamefont {Ji}\ \emph {et~al.}(2024)\citenamefont {Ji},
  \citenamefont {Yu}, \citenamefont {Xu},\ and\ \citenamefont {Xiang}}]{FQFE}%
  \BibitemOpen
  \bibfield  {author} {\bibinfo {author} {\bibfnamefont {J.}~\bibnamefont
  {Ji}}, \bibinfo {author} {\bibfnamefont {G.}~\bibnamefont {Yu}}, \bibinfo
  {author} {\bibfnamefont {C.}~\bibnamefont {Xu}},\ and\ \bibinfo {author}
  {\bibfnamefont {H.}~\bibnamefont {Xiang}},\ }\bibfield  {title} {\bibinfo
  {title} {Fractional quantum ferroelectricity},\ }\href@noop {} {\bibfield
  {journal} {\bibinfo  {journal} {Nat. Commun.}\ }\textbf {\bibinfo {volume}
  {15}},\ \bibinfo {pages} {135} (\bibinfo {year} {2024})}\BibitemShut
  {NoStop}%
\bibitem [{\citenamefont {Pang}\ and\ \citenamefont
  {He}(2025{\natexlab{a}})}]{pang2025generalized}%
  \BibitemOpen
  \bibfield  {author} {\bibinfo {author} {\bibfnamefont {H.}~\bibnamefont
  {Pang}}\ and\ \bibinfo {author} {\bibfnamefont {L.}~\bibnamefont {He}},\
  }\bibfield  {title} {\bibinfo {title} {Generalized neumann’s principle as a
  unified framework for fractional quantum and conventional ferroelectricity},\
  }\href@noop {} {\bibfield  {journal} {\bibinfo  {journal} {Phys. Rev. Lett.}\
  }\textbf {\bibinfo {volume} {135}},\ \bibinfo {pages} {116402} (\bibinfo
  {year} {2025}{\natexlab{a}})}\BibitemShut {NoStop}%
\bibitem [{\citenamefont {Pang}\ and\ \citenamefont
  {He}(2025{\natexlab{b}})}]{pang2025quantized}%
  \BibitemOpen
  \bibfield  {author} {\bibinfo {author} {\bibfnamefont {H.}~\bibnamefont
  {Pang}}\ and\ \bibinfo {author} {\bibfnamefont {L.}~\bibnamefont {He}},\
  }\bibfield  {title} {\bibinfo {title} {Quantized polarization redefines polar
  interfaces},\ }\href@noop {} {\bibfield  {journal} {\bibinfo  {journal}
  {arXiv preprint arXiv:2511.18697}\ } (\bibinfo {year}
  {2025}{\natexlab{b}})}\BibitemShut {NoStop}%
\bibitem [{\citenamefont {Pang}\ and\ \citenamefont
  {He}(2026)}]{pang2026insulator}%
  \BibitemOpen
  \bibfield  {author} {\bibinfo {author} {\bibfnamefont {H.}~\bibnamefont
  {Pang}}\ and\ \bibinfo {author} {\bibfnamefont {L.}~\bibnamefont {He}},\
  }\bibfield  {title} {\bibinfo {title} {Insulator-to-metal transitions driven
  by quantized formal polarization mismatch},\ }\href@noop {} {\bibfield
  {journal} {\bibinfo  {journal} {arXiv preprint arXiv:2604.01530}\ } (\bibinfo
  {year} {2026})}\BibitemShut {NoStop}%
\bibitem [{SM()}]{SM}%
  \BibitemOpen
  \href@noop {} {}\bibinfo {note} {See Supplemental Material for response
  formulas for second-harmonic generation and shift current, computational
  details, symmetry-allowed second-order tensor components, and supplemental
  results for representative materials, including InPS$_3$, InNbBr$_6$, and
  CdBiO$_3$. The Supplemental Material includes
  Refs.~\cite{sipe,rashkeev,young2012first,fregoso2017quantitative,sturman2021photovoltaic,li2016large,
  chen2010systematically,lin2024abinitio,lin2021strategy,hse,lin2020accuracy,Hamann2013ONCV,Scherpelz2016SG15,
  jin2023,haastrup2018computational,gjerding2021recent,pang2025generalized,
  chen1977ionic,hsu1990ab,lin1999mechanism,trinquet2024second}.}\BibitemShut
  {Stop}%
\bibitem [{\citenamefont {Jin}\ and\ \citenamefont
  {He}(2024)}]{jin2024peculiar}%
  \BibitemOpen
  \bibfield  {author} {\bibinfo {author} {\bibfnamefont {G.}~\bibnamefont
  {Jin}}\ and\ \bibinfo {author} {\bibfnamefont {L.}~\bibnamefont {He}},\
  }\bibfield  {title} {\bibinfo {title} {Peculiar band geometry induced giant
  shift current in ferroelectric snte monolayer},\ }\href@noop {} {\bibfield
  {journal} {\bibinfo  {journal} {npj Comput. Mater.}\ }\textbf {\bibinfo
  {volume} {10}},\ \bibinfo {pages} {23} (\bibinfo {year} {2024})}\BibitemShut
  {NoStop}%
\bibitem [{\citenamefont {Young}\ and\ \citenamefont
  {Rappe}(2012)}]{young2012first}%
  \BibitemOpen
  \bibfield  {author} {\bibinfo {author} {\bibfnamefont {S.~M.}\ \bibnamefont
  {Young}}\ and\ \bibinfo {author} {\bibfnamefont {A.~M.}\ \bibnamefont
  {Rappe}},\ }\bibfield  {title} {\bibinfo {title} {First principles
  calculation of the shift current photovoltaic effect in ferroelectrics},\
  }\href@noop {} {\bibfield  {journal} {\bibinfo  {journal} {Phys. Rev. Lett.}\
  }\textbf {\bibinfo {volume} {109}},\ \bibinfo {pages} {116601} (\bibinfo
  {year} {2012})}\BibitemShut {NoStop}%
\bibitem [{\citenamefont {Huang}\ \emph {et~al.}(2024)\citenamefont {Huang},
  \citenamefont {Xiao}, \citenamefont {Xia}, \citenamefont {Chen},\ and\
  \citenamefont {Zhai}}]{huang2024second}%
  \BibitemOpen
  \bibfield  {author} {\bibinfo {author} {\bibfnamefont {W.}~\bibnamefont
  {Huang}}, \bibinfo {author} {\bibfnamefont {Y.}~\bibnamefont {Xiao}},
  \bibinfo {author} {\bibfnamefont {F.}~\bibnamefont {Xia}}, \bibinfo {author}
  {\bibfnamefont {X.}~\bibnamefont {Chen}},\ and\ \bibinfo {author}
  {\bibfnamefont {T.}~\bibnamefont {Zhai}},\ }\bibfield  {title} {\bibinfo
  {title} {Second harmonic generation control in 2d layered materials: status
  and outlook},\ }\href@noop {} {\bibfield  {journal} {\bibinfo  {journal}
  {Adv. Funct. Mater.}\ }\textbf {\bibinfo {volume} {34}},\ \bibinfo {pages}
  {2310726} (\bibinfo {year} {2024})}\BibitemShut {NoStop}%
\bibitem [{\citenamefont {Zhang}\ \emph {et~al.}(2020)\citenamefont {Zhang},
  \citenamefont {Zhao}, \citenamefont {Yu}, \citenamefont {Yang},\ and\
  \citenamefont {Liu}}]{zhang2020second}%
  \BibitemOpen
  \bibfield  {author} {\bibinfo {author} {\bibfnamefont {J.}~\bibnamefont
  {Zhang}}, \bibinfo {author} {\bibfnamefont {W.}~\bibnamefont {Zhao}},
  \bibinfo {author} {\bibfnamefont {P.}~\bibnamefont {Yu}}, \bibinfo {author}
  {\bibfnamefont {G.}~\bibnamefont {Yang}},\ and\ \bibinfo {author}
  {\bibfnamefont {Z.}~\bibnamefont {Liu}},\ }\bibfield  {title} {\bibinfo
  {title} {Second harmonic generation in 2d layered materials},\ }\href@noop {}
  {\bibfield  {journal} {\bibinfo  {journal} {2D Mater.}\ }\textbf {\bibinfo
  {volume} {7}},\ \bibinfo {pages} {042002} (\bibinfo {year}
  {2020})}\BibitemShut {NoStop}%
\bibitem [{\citenamefont {Haastrup}\ \emph {et~al.}(2018)\citenamefont
  {Haastrup}, \citenamefont {Strange}, \citenamefont {Pandey}, \citenamefont
  {Deilmann}, \citenamefont {Schmidt}, \citenamefont {Hinsche}, \citenamefont
  {Gjerding}, \citenamefont {Torelli}, \citenamefont {Larsen}, \citenamefont
  {Riis-Jensen} \emph {et~al.}}]{haastrup2018computational}%
  \BibitemOpen
  \bibfield  {author} {\bibinfo {author} {\bibfnamefont {S.}~\bibnamefont
  {Haastrup}}, \bibinfo {author} {\bibfnamefont {M.}~\bibnamefont {Strange}},
  \bibinfo {author} {\bibfnamefont {M.}~\bibnamefont {Pandey}}, \bibinfo
  {author} {\bibfnamefont {T.}~\bibnamefont {Deilmann}}, \bibinfo {author}
  {\bibfnamefont {P.~S.}\ \bibnamefont {Schmidt}}, \bibinfo {author}
  {\bibfnamefont {N.~F.}\ \bibnamefont {Hinsche}}, \bibinfo {author}
  {\bibfnamefont {M.~N.}\ \bibnamefont {Gjerding}}, \bibinfo {author}
  {\bibfnamefont {D.}~\bibnamefont {Torelli}}, \bibinfo {author} {\bibfnamefont
  {P.~M.}\ \bibnamefont {Larsen}}, \bibinfo {author} {\bibfnamefont {A.~C.}\
  \bibnamefont {Riis-Jensen}}, \emph {et~al.},\ }\bibfield  {title} {\bibinfo
  {title} {The computational 2d materials database: high-throughput modeling
  and discovery of atomically thin crystals},\ }\href@noop {} {\bibfield
  {journal} {\bibinfo  {journal} {2D Mater.}\ }\textbf {\bibinfo {volume}
  {5}},\ \bibinfo {pages} {042002} (\bibinfo {year} {2018})}\BibitemShut
  {NoStop}%
\bibitem [{\citenamefont {Gjerding}\ \emph {et~al.}(2021)\citenamefont
  {Gjerding}, \citenamefont {Taghizadeh}, \citenamefont {Rasmussen},
  \citenamefont {Ali}, \citenamefont {Bertoldo}, \citenamefont {Deilmann},
  \citenamefont {Kn{\o}sgaard}, \citenamefont {Kruse}, \citenamefont {Larsen},
  \citenamefont {Manti} \emph {et~al.}}]{gjerding2021recent}%
  \BibitemOpen
  \bibfield  {author} {\bibinfo {author} {\bibfnamefont {M.~N.}\ \bibnamefont
  {Gjerding}}, \bibinfo {author} {\bibfnamefont {A.}~\bibnamefont
  {Taghizadeh}}, \bibinfo {author} {\bibfnamefont {A.}~\bibnamefont
  {Rasmussen}}, \bibinfo {author} {\bibfnamefont {S.}~\bibnamefont {Ali}},
  \bibinfo {author} {\bibfnamefont {F.}~\bibnamefont {Bertoldo}}, \bibinfo
  {author} {\bibfnamefont {T.}~\bibnamefont {Deilmann}}, \bibinfo {author}
  {\bibfnamefont {N.~R.}\ \bibnamefont {Kn{\o}sgaard}}, \bibinfo {author}
  {\bibfnamefont {M.}~\bibnamefont {Kruse}}, \bibinfo {author} {\bibfnamefont
  {A.~H.}\ \bibnamefont {Larsen}}, \bibinfo {author} {\bibfnamefont
  {S.}~\bibnamefont {Manti}}, \emph {et~al.},\ }\bibfield  {title} {\bibinfo
  {title} {Recent progress of the computational 2d materials database (c2db)},\
  }\href@noop {} {\bibfield  {journal} {\bibinfo  {journal} {2D Mater.}\
  }\textbf {\bibinfo {volume} {8}},\ \bibinfo {pages} {044002} (\bibinfo {year}
  {2021})}\BibitemShut {NoStop}%
\bibitem [{\citenamefont {Heyd}\ \emph {et~al.}(2003)\citenamefont {Heyd},
  \citenamefont {Scuseria},\ and\ \citenamefont {Ernzerhof}}]{hse}%
  \BibitemOpen
  \bibfield  {author} {\bibinfo {author} {\bibfnamefont {J.}~\bibnamefont
  {Heyd}}, \bibinfo {author} {\bibfnamefont {G.~E.}\ \bibnamefont {Scuseria}},\
  and\ \bibinfo {author} {\bibfnamefont {M.}~\bibnamefont {Ernzerhof}},\
  }\bibfield  {title} {\bibinfo {title} {Hybrid functionals based on a screened
  coulomb potential},\ }\href@noop {} {\bibfield  {journal} {\bibinfo
  {journal} {J. Chem. Phys}\ }\textbf {\bibinfo {volume} {118}},\ \bibinfo
  {pages} {8207} (\bibinfo {year} {2003})}\BibitemShut {NoStop}%
\bibitem [{\citenamefont {Li}\ \emph {et~al.}(2016)\citenamefont {Li},
  \citenamefont {Liu}, \citenamefont {Chen}, \citenamefont {Lin}, \citenamefont
  {Ren}, \citenamefont {Lin}, \citenamefont {Yang},\ and\ \citenamefont
  {He}}]{li2016large}%
  \BibitemOpen
  \bibfield  {author} {\bibinfo {author} {\bibfnamefont {P.}~\bibnamefont
  {Li}}, \bibinfo {author} {\bibfnamefont {X.}~\bibnamefont {Liu}}, \bibinfo
  {author} {\bibfnamefont {M.}~\bibnamefont {Chen}}, \bibinfo {author}
  {\bibfnamefont {P.}~\bibnamefont {Lin}}, \bibinfo {author} {\bibfnamefont
  {X.}~\bibnamefont {Ren}}, \bibinfo {author} {\bibfnamefont {L.}~\bibnamefont
  {Lin}}, \bibinfo {author} {\bibfnamefont {C.}~\bibnamefont {Yang}},\ and\
  \bibinfo {author} {\bibfnamefont {L.}~\bibnamefont {He}},\ }\bibfield
  {title} {\bibinfo {title} {Large-scale {ab initio} simulations based on
  systematically improvable atomic basis},\ }\href
  {https://doi.org/10.1016/j.commatsci.2015.07.004} {\bibfield  {journal}
  {\bibinfo  {journal} {Comput. Mater. Sci.}\ }\textbf {\bibinfo {volume}
  {112}},\ \bibinfo {pages} {503} (\bibinfo {year} {2016})}\BibitemShut
  {NoStop}%
\bibitem [{\citenamefont {Chen}\ \emph {et~al.}(2010)\citenamefont {Chen},
  \citenamefont {Guo},\ and\ \citenamefont {He}}]{chen2010systematically}%
  \BibitemOpen
  \bibfield  {author} {\bibinfo {author} {\bibfnamefont {M.}~\bibnamefont
  {Chen}}, \bibinfo {author} {\bibfnamefont {G.-C.}\ \bibnamefont {Guo}},\ and\
  \bibinfo {author} {\bibfnamefont {L.}~\bibnamefont {He}},\ }\bibfield
  {title} {\bibinfo {title} {Systematically improvable optimized atomic basis
  sets for {ab initio} calculations},\ }\href
  {https://doi.org/10.1088/0953-8984/22/44/445501} {\bibfield  {journal}
  {\bibinfo  {journal} {J. Phys.: Condens. Matter}\ }\textbf {\bibinfo {volume}
  {22}},\ \bibinfo {pages} {445501} (\bibinfo {year} {2010})}\BibitemShut
  {NoStop}%
\bibitem [{\citenamefont {Lin}\ \emph {et~al.}(2024)\citenamefont {Lin},
  \citenamefont {Ren}, \citenamefont {Liu},\ and\ \citenamefont
  {He}}]{lin2024abinitio}%
  \BibitemOpen
  \bibfield  {author} {\bibinfo {author} {\bibfnamefont {P.}~\bibnamefont
  {Lin}}, \bibinfo {author} {\bibfnamefont {X.}~\bibnamefont {Ren}}, \bibinfo
  {author} {\bibfnamefont {X.}~\bibnamefont {Liu}},\ and\ \bibinfo {author}
  {\bibfnamefont {L.}~\bibnamefont {He}},\ }\bibfield  {title} {\bibinfo
  {title} {{Ab initio} electronic structure calculations based on numerical
  atomic orbitals: Basic fomalisms and recent progresses},\ }\href
  {https://doi.org/10.1002/wcms.1687} {\bibfield  {journal} {\bibinfo
  {journal} {WIREs Comput. Mol. Sci.}\ }\textbf {\bibinfo {volume} {14}},\
  \bibinfo {pages} {e1687} (\bibinfo {year} {2024})}\BibitemShut {NoStop}%
\bibitem [{\citenamefont {Lin}\ \emph {et~al.}(2021)\citenamefont {Lin},
  \citenamefont {Ren},\ and\ \citenamefont {He}}]{lin2021strategy}%
  \BibitemOpen
  \bibfield  {author} {\bibinfo {author} {\bibfnamefont {P.}~\bibnamefont
  {Lin}}, \bibinfo {author} {\bibfnamefont {X.}~\bibnamefont {Ren}},\ and\
  \bibinfo {author} {\bibfnamefont {L.}~\bibnamefont {He}},\ }\bibfield
  {title} {\bibinfo {title} {Strategy for constructing compact numerical atomic
  orbital basis sets by incorporating the gradients of reference
  wavefunctions},\ }\href {https://doi.org/10.1103/PhysRevB.103.235131}
  {\bibfield  {journal} {\bibinfo  {journal} {Phys. Rev. B}\ }\textbf {\bibinfo
  {volume} {103}},\ \bibinfo {pages} {235131} (\bibinfo {year}
  {2021})}\BibitemShut {NoStop}%
\bibitem [{\citenamefont {Lin}\ \emph {et~al.}(2020)\citenamefont {Lin},
  \citenamefont {Ren},\ and\ \citenamefont {He}}]{lin2020accuracy}%
  \BibitemOpen
  \bibfield  {author} {\bibinfo {author} {\bibfnamefont {P.}~\bibnamefont
  {Lin}}, \bibinfo {author} {\bibfnamefont {X.}~\bibnamefont {Ren}},\ and\
  \bibinfo {author} {\bibfnamefont {L.}~\bibnamefont {He}},\ }\bibfield
  {title} {\bibinfo {title} {Accuracy of localized resolution of the identity
  in periodic hybrid functional calculations with numerical atomic orbitals},\
  }\href {https://doi.org/10.1021/acs.jpclett.0c00481} {\bibfield  {journal}
  {\bibinfo  {journal} {J. Phys. Chem. Lett.}\ }\textbf {\bibinfo {volume}
  {11}},\ \bibinfo {pages} {3082} (\bibinfo {year} {2020})}\BibitemShut
  {NoStop}%
\bibitem [{\citenamefont {Jin}\ \emph {et~al.}(2023)\citenamefont {Jin},
  \citenamefont {Pang}, \citenamefont {Ji}, \citenamefont {Dai},\ and\
  \citenamefont {He}}]{jin2023}%
  \BibitemOpen
  \bibfield  {author} {\bibinfo {author} {\bibfnamefont {G.}~\bibnamefont
  {Jin}}, \bibinfo {author} {\bibfnamefont {H.}~\bibnamefont {Pang}}, \bibinfo
  {author} {\bibfnamefont {Y.}~\bibnamefont {Ji}}, \bibinfo {author}
  {\bibfnamefont {Z.}~\bibnamefont {Dai}},\ and\ \bibinfo {author}
  {\bibfnamefont {L.}~\bibnamefont {He}},\ }\bibfield  {title} {\bibinfo
  {title} {Pyatb: An efficient python package for electronic structure
  calculations using ab initio tight-binding model},\ }\href
  {https://doi.org/https://doi.org/10.1016/j.cpc.2023.108844} {\bibfield
  {journal} {\bibinfo  {journal} {Comput. Phys. Commun.}\ }\textbf {\bibinfo
  {volume} {291}},\ \bibinfo {pages} {108844} (\bibinfo {year}
  {2023})}\BibitemShut {NoStop}%
\bibitem [{\citenamefont {Wang}\ and\ \citenamefont
  {Qian}(2017)}]{wang2017giant}%
  \BibitemOpen
  \bibfield  {author} {\bibinfo {author} {\bibfnamefont {H.}~\bibnamefont
  {Wang}}\ and\ \bibinfo {author} {\bibfnamefont {X.}~\bibnamefont {Qian}},\
  }\bibfield  {title} {\bibinfo {title} {Giant optical second harmonic
  generation in two-dimensional multiferroics},\ }\href@noop {} {\bibfield
  {journal} {\bibinfo  {journal} {Nano Lett.}\ }\textbf {\bibinfo {volume}
  {17}},\ \bibinfo {pages} {5027} (\bibinfo {year} {2017})}\BibitemShut
  {NoStop}%
\bibitem [{\citenamefont {Hamann}(2013)}]{Hamann2013ONCV}%
  \BibitemOpen
  \bibfield  {author} {\bibinfo {author} {\bibfnamefont {D.~R.}\ \bibnamefont
  {Hamann}},\ }\bibfield  {title} {\bibinfo {title} {Optimized norm-conserving
  vanderbilt pseudopotentials},\ }\href
  {https://doi.org/10.1103/PhysRevB.88.085117} {\bibfield  {journal} {\bibinfo
  {journal} {Phys. Rev. B}\ }\textbf {\bibinfo {volume} {88}},\ \bibinfo
  {pages} {085117} (\bibinfo {year} {2013})}\BibitemShut {NoStop}%
\bibitem [{\citenamefont {Scherpelz}\ \emph {et~al.}(2016)\citenamefont
  {Scherpelz}, \citenamefont {Govoni}, \citenamefont {Hamada},\ and\
  \citenamefont {Galli}}]{Scherpelz2016SG15}%
  \BibitemOpen
  \bibfield  {author} {\bibinfo {author} {\bibfnamefont {P.}~\bibnamefont
  {Scherpelz}}, \bibinfo {author} {\bibfnamefont {M.}~\bibnamefont {Govoni}},
  \bibinfo {author} {\bibfnamefont {I.}~\bibnamefont {Hamada}},\ and\ \bibinfo
  {author} {\bibfnamefont {G.}~\bibnamefont {Galli}},\ }\bibfield  {title}
  {\bibinfo {title} {Implementation and validation of fully relativistic gw
  calculations: Spin–orbit coupling in molecules, nanocrystals, and solids},\
  }\href {https://doi.org/10.1021/acs.jctc.6b00114} {\bibfield  {journal}
  {\bibinfo  {journal} {J. Chem. Theory Comput.}\ }\textbf {\bibinfo {volume}
  {12}},\ \bibinfo {pages} {3523} (\bibinfo {year} {2016})}\BibitemShut
  {NoStop}%
\bibitem [{\citenamefont {Chen}(1977)}]{chen1977ionic}%
  \BibitemOpen
  \bibfield  {author} {\bibinfo {author} {\bibfnamefont {C.~T.}\ \bibnamefont
  {Chen}},\ }\bibfield  {title} {\bibinfo {title} {An ionic grouping theory of
  the electro-optical and nonlinear optical effects of crystals. {III}. a
  theoretical calculation of the electro-optical and optical second-harmonic
  coefficients of {LiNbO$_3$}, {LiTaO$_3$}, {KNbO$_3$}, and {BNN} crystals
  based on a deformed oxygen-octahedra ionic grouping model},\ }\href
  {https://doi.org/10.7498/APS.26.486} {\bibfield  {journal} {\bibinfo
  {journal} {Acta Phys. Sin.}\ }\textbf {\bibinfo {volume} {26}},\ \bibinfo
  {pages} {486} (\bibinfo {year} {1977})}\BibitemShut {NoStop}%
\bibitem [{\citenamefont {Hsu}\ \emph {et~al.}(1990)\citenamefont {Hsu},
  \citenamefont {Kasowski}, \citenamefont {Ma},\ and\ \citenamefont
  {Chui}}]{hsu1990ab}%
  \BibitemOpen
  \bibfield  {author} {\bibinfo {author} {\bibfnamefont {W.~Y.}\ \bibnamefont
  {Hsu}}, \bibinfo {author} {\bibfnamefont {R.~V.}\ \bibnamefont {Kasowski}},
  \bibinfo {author} {\bibfnamefont {H.}~\bibnamefont {Ma}},\ and\ \bibinfo
  {author} {\bibfnamefont {S.~T.}\ \bibnamefont {Chui}},\ }\bibfield  {title}
  {\bibinfo {title} {{Ab initio} computation of the linear and nonlinear
  optical properties of {LiNbO$_3$}, and {KTP}},\ }in\ \href@noop {} {\emph
  {\bibinfo {booktitle} {International Quantum Electronics Conference}}},\
  \bibinfo {series} {OSA Technical Digest}, Vol.~\bibinfo {volume} {8},\
  \bibinfo {editor} {edited by\ \bibinfo {editor} {\bibfnamefont
  {A.}~\bibnamefont {Owyoung}}, \bibinfo {editor} {\bibfnamefont
  {C.}~\bibnamefont {Shank}}, \bibinfo {editor} {\bibfnamefont
  {S.}~\bibnamefont {Chu}},\ and\ \bibinfo {editor} {\bibfnamefont
  {E.}~\bibnamefont {Ippen}}}\ (\bibinfo  {publisher} {Optica Publishing
  Group},\ \bibinfo {year} {1990})\ p.\ \bibinfo {pages} {QWD27}\BibitemShut
  {NoStop}%
\bibitem [{\citenamefont {Lin}\ \emph {et~al.}(1999)\citenamefont {Lin},
  \citenamefont {Lee}, \citenamefont {Liu}, \citenamefont {Chen},\ and\
  \citenamefont {Pickard}}]{lin1999mechanism}%
  \BibitemOpen
  \bibfield  {author} {\bibinfo {author} {\bibfnamefont {J.}~\bibnamefont
  {Lin}}, \bibinfo {author} {\bibfnamefont {M.-H.}\ \bibnamefont {Lee}},
  \bibinfo {author} {\bibfnamefont {Z.-P.}\ \bibnamefont {Liu}}, \bibinfo
  {author} {\bibfnamefont {C.}~\bibnamefont {Chen}},\ and\ \bibinfo {author}
  {\bibfnamefont {C.~J.}\ \bibnamefont {Pickard}},\ }\bibfield  {title}
  {\bibinfo {title} {Mechanism for linear and nonlinear optical effects in
  {$\beta$-{BaB$_2$O$_4$}} crystals},\ }\href
  {https://doi.org/10.1103/PhysRevB.60.13380} {\bibfield  {journal} {\bibinfo
  {journal} {Phys. Rev. B}\ }\textbf {\bibinfo {volume} {60}},\ \bibinfo
  {pages} {13380} (\bibinfo {year} {1999})}\BibitemShut {NoStop}%
\bibitem [{\citenamefont {Trinquet}\ \emph {et~al.}(2024)\citenamefont
  {Trinquet}, \citenamefont {Naccarato}, \citenamefont {Brunin}, \citenamefont
  {Petretto}, \citenamefont {Wirtz}, \citenamefont {Hautier},\ and\
  \citenamefont {Rignanese}}]{trinquet2024second}%
  \BibitemOpen
  \bibfield  {author} {\bibinfo {author} {\bibfnamefont {V.}~\bibnamefont
  {Trinquet}}, \bibinfo {author} {\bibfnamefont {F.}~\bibnamefont {Naccarato}},
  \bibinfo {author} {\bibfnamefont {G.}~\bibnamefont {Brunin}}, \bibinfo
  {author} {\bibfnamefont {G.}~\bibnamefont {Petretto}}, \bibinfo {author}
  {\bibfnamefont {L.}~\bibnamefont {Wirtz}}, \bibinfo {author} {\bibfnamefont
  {G.}~\bibnamefont {Hautier}},\ and\ \bibinfo {author} {\bibfnamefont {G.-M.}\
  \bibnamefont {Rignanese}},\ }\bibfield  {title} {\bibinfo {title}
  {Second-harmonic generation tensors from high-throughput density-functional
  perturbation theory},\ }\href@noop {} {\bibfield  {journal} {\bibinfo
  {journal} {Sci. Data}\ }\textbf {\bibinfo {volume} {11}},\ \bibinfo {pages}
  {757} (\bibinfo {year} {2024})}\BibitemShut {NoStop}%
\end{thebibliography}
	

\end{document}